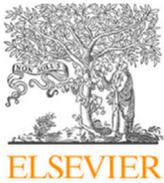
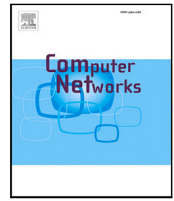

Survey paper

# Intent-driven autonomous network and service management in future cellular networks: A structured literature review

Kashif Mehmood *, Katina Kralevska, David Palma

*Department of Information Security and Communication Technology, Norwegian University of Science and Technology (NTNU), Trondheim, Norway*

ABSTRACT

Intent-driven networks are an essential stepping stone in the evolution of network and service management towards a truly autonomous paradigm. User centric intents provide an abstracted means of impacting the design, provisioning, deployment and assurance of network infrastructure and services with the help of service level agreements and minimum network capability exposure. The concept of Intent Based Networking (IBN) poses several challenges in terms of the contextual definition of intents, role of different stakeholders, and a generalized architecture. In this review, we provide a comprehensive analysis of the state-of-the-art in IBN including the intent description models, intent lifecycle management, significance of IBN and a generalized architectural framework along with challenges and prospects for IBN in future cellular networks. An analytical study is performed on the data collected from relevant studies primarily focusing on the interworking of IBN with softwarized networking based on NFV/SDN infrastructures. Critical functions required in the IBN management and service model design are explored with different abstract modeling techniques and a converged architectural framework is proposed. The key findings include: (1) benefits and role of IBN in autonomous networking, (2) improvements needed to integrate intents as fundamental policies for service modeling and network management, (3) need for appropriate representation models for intents in domain agnostic abstract manner, and (4) need to include learning as a fundamental function in autonomous networks. These observations provide the basis for in-depth investigation and standardization efforts for IBN as a fundamental network management paradigm in beyond 5G cellular networks.

## 1. Introduction

Traditional network and service management consists of several modules comprised mainly of human experts as well as automated scripts and functionalities to monitor, tune and decide configuration requirements for different services [1]. The level of autonomy for the provisioning and management of these networks and their functions are dictated by the human experts and the guidelines provided by the network service providers, the available resources and the required resources for user service provision. This approach met the needs of the previous generation networks quite efficiently. However, the rapid need for evolution and the drastic changes made it difficult to keep up and train the experts with the required expertise in a timely manner. The associated challenges that require addressing include reduction in deployment cost, retraining of network administrators, decision making and the human inability to adapt. *Autonomous network management* [2] has attracted significant attention from the networking community as a possible solution to address these challenges and improve the process of network and service provisioning for a diverse set of user and service provider requirements.

Intent-driven networking is poised to provide a design and performance enhancement of communication networks. A higher level intervention from the users is regarded as an *intent* [2]. The objective is to provide guidance in the form of network understandable input from the users. The required information from user to interact with the network through intents does not encompass low-level or configuration level information. An autonomous network is able to understand this intent from the users and pass it through the layers, extracting required information for different network nodes, finally providing a configuration for the involved network functions.

The motivation for integrating intent-based decision making in future networks is inspired by the possibility of abstract network management and resource provisioning, driven by intents from the users and service providers in the network. The design of network configuration and adaptability according to the preferences of the subscribers presents an interesting direction for network infrastructure providers in future networks. This abstracted policy input from the network stakeholders enables the necessary agility and adaptability for the service

| **Acronyms** | |
|---|---|
| 5G | Fifth generation |
| AI | Artificial intelligence |
| API | Application programming interface |
| CFCS | Customer-facing communication service |
| CSC | Communication service customer |
| CSI | Communication service instances |
| CSMF | Communication service management function |
| CSP | Communication service provider |
| e2e | End-to-end |
| ENI | Experiential network intelligence |
| ETSI | European telecommunications standards institute |
| GAN | Generative adversarial network |
| GSMA | GSM association |
| GST | Generic slicing template |
| IBN | Intent based networking |
| IETF | Internet engineering task force |
| ITU | International telecommunications union |
| KPI | Key performance indicator |
| LCM | Lifecycle management |
| ML | Machine learning |
| MLaaS | Machine learning as a service |
| NBI | Northbound interface |
| NFV | Network function virtualization |
| NLP | Natural language processing |
| ONOS | Open network operating system |
| OSM | Open source MANO |
| QoS | Quality of service |
| REST | Representational state transfer |
| RFCS | Resource-facing communication service |
| RNN | Recurrent neural network |
| SBA | Service-based architecture |
| SBI | Southbound interface |
| SD | Service descriptor |
| SDN | Software defined networking |
| SFC | Service function chains |
| SIG | Softgoal inter-dependency graph |
| SLA | Service level agreement |
| SP | Service provider |
| TOSCA | Topology and orchestration specification for cloud applications |
| VNF | Virtualized network function |
| YANG | Yet another next generation |
| ZSM | Zero-touch network and service management |

and infrastructure management. However, the challenge of understanding these user and service provider intents remains an open issue in the networking community, along with the feasibility, verification and provision of resources.

Network and service management has been focused towards a narrow and micro-managed paradigm involving low-level device and feature planning [3]. This approach has limitations for larger networks and higher device densities, making it infeasible to scale. Initially, International Telecommunications Union (ITU) Telecommunications Management Network (TMN) [1] promised a *hierarchical management model*, introducing the role of business and device objectives for network configurations. This model provided the much needed distribution of management functions across various layers and simplified the process. A level of control required to configure devices according to the business and network infrastructure demands was also provided. The concept also inspired the need for a higher level control of network entities in an efficient and more effective manner.

*1.1. Standardization and related surveys*

The autonomic networking initiative is driven by European Telecommunications Standards Institute (ETSI) and Internet Engineering Task Force (IETF) with relevant working groups namely Zero-touch Network and Service Management (ZSM) [4], Experiential Network Intelligence (ENI) [5], and Autonomic Networking Integrated Model and Approach (ANIMA) [6]. ZSM proposed an end-to-end (e2e) automation framework for network, service management, and sustainable development of a diverse set of services. This approach fulfills the shortcomings of the traditional human centric network management along with the need to support software-defined, flexible and service-based architectures. Moreover, an autonomous network framework ensures self-configuring, healing, monitoring and deployment of network functions. The architecture, potential use-cases and requirements are highlighted through the ZSM initiative.

The ENI [7–9] initiative focuses on achieving a cognitive network management architecture using machine learning and context-aware policies to address heterogeneous service requirements based on the user needs, network conditions and the business objectives of service providers. ENI encompasses Artificial Intelligence (AI) capabilities in a closed-loop control mechanism to improve performance of network orchestration and resource management functionalities. The intelligence is offered to the emerging technologies like network slicing, Software Defined Networking (SDN) and Network Function Virtualization (NFV) for enhancement of their operational and maintenance lifecycles.

IBN provides an inherently intelligent framework for representing user and service provider expectations in terms of abstract policies referred to as intents [13]. IBN includes the necessary functionality to understand and interpret the intents from different stakeholders and convert them into well-defined service requirements. The service level information is extracted and verified for possible conflicts to other intents in consultation with the Service Level Agreements (SLAs). The service-specific configuration templates are generated after understanding the context provided by the intents through the processing, mapping and conflict resolution functions in the intent lifecycle. For example, this step can include the provisioning of Service Function Chains (SFC) required to satisfy the service and the user needs eventually in an NFV environment. The advantage of using an intent-driven approach lies in the abstractness that can be achieved with the provided intents and letting the intent translation, evaluation and verification modules decide about the actual deployment policy. The networking management group under IETF [14] is working on the standardization and documentation of the concepts behind IBN and the possible migration paths from the traditional networks.

The motivation of utilizing IBN is inspired by the development and management of network controller and service based architecture in various domains including but not restricted to cellular, data center, and enterprise environments. In a related work, Rafiq et al. [15] analyze the load distribution in data center environments using a novel IBN manager providing increased utilization of network resources in SDN. In another study [16], Machine Learning (ML) models are utilized for intent representation and time series forecasting of required resources in a data center environment. This is achieved using a combination of convolutional, recurrent neural networks to assure the performance of different network services. Furthermore, enterprise networks are the focus of [17], where an intent-based software-defined wide area network orchestrator is proposed for network slicing. The proposed orchestrator provides a flexible solution for domain-agnostic e2e management of network slices with validation through an ingeniously designed testbed.

Focused investigations on intent-driven networking as an enabler for service-based architecture, have been reported but the scope is limited and few potential advantages and challenges of IBN are provided.

**Table 1**
A comparison of related surveys and their scope.

| Scope | Surveys | | | |
|---|---|---|---|---|
| | Yiming et al. [10] (2020) | Lei et al.[11] (2020) | Engin et al. [12] (2020) | Our work (2021) |
| Intent semantics | ✓ | ✓ | ✓ | ✓ |
| Intent standardization | ✗ | ✓ | ✓ | ✓ |
| Intent lifecycle and processing | ✓ | ✗ | ✗ | ✓ |
| IBN orchestration | ✗ | ✓ | ✗ | ✓ |
| Intent domains and use cases | ✗ | ✓ | ✗ | ✓ |
| IBN management architecture | ✓ | ✗ | ✗ | ✓ |
| Communication service and network management | ✗ | ✗ | ✗ | ✓ |

**Table 2**
Literature search keywords and constructed phrases.

| Network management | Service management | IBN and autonomous design | Search phrases |
|---|---|---|---|
| networking, network management, network orchestration, network virtualization, networking slicing, software defined, SDN | services, service orchestration, service management service model, service design | intent, intent based, intent driven, intent configuration, autonomous, intelligence, automation | – ("intent based networking"), ("intent driven networking"), ("intent configuration"), ("intent service model")<br>– (("network management" OR "network orchestration" OR "network virtualization" OR "network slicing") AND "intent")<br>– (("service management" OR "service orchestration") AND "intent")<br>– (("SDN" OR "software defined") AND "intent")<br>– (("autonomous network" OR "network intelligence") AND "intent") |

Furthermore, a discussion of potential tradeoffs involved in the process of intent provisioning and mapping onto actual network configurations for different services is lacking.

The survey by Wei et al. [10] explains different aspects of IBN with regards to the possibilities in future networks including Sixth Generation (6G), different potential markets as well as industrial players involved in the development of IBN architectures. The study explains the basics of an IBN architecture, key technologies and the networking perspective in regards to programmed virtualized network functions in potential 6G networks.

A list of industrial implementations of IBN is provided along with a focus on intent-driven network operation and maintenance. The prominent ones are from Cisco, Huawei, Juniper and Spruce Network, focused towards a centralized design, involving data centers and a converged cloud infrastructure. However, a complete framework visualizing different segments in future networks is missing, as well as a converged architecture and a discussion on different standardization efforts.

In another survey, Pang et al. [11] provide a comprehensive overview of the IBN architecture and different standardization milestones. Potential application scenarios are also highlighted followed by several challenges in the adoption of IBN in deployed and future networks. However, the architecture description is particularly vague and lacks the appropriate connection or migration paths from the traditional networking technologies.

A recent survey by Zeydan and Turk [12] provides an overview of intent-based techniques from a network management and orchestration perspective. The study summarizes different standardization efforts related to intent-based management and provides a broad overview of intent semantics and lifecycle.

The surveys [10–12] lack the consideration and a deeper analysis of the connection between subscriber and service provider intents. Moreover, potential tradeoffs involved in the processing, dissemination and application of intents in the network architecture are not explored. The scope of relevant survey articles is listed in Table 1 highlighting the need for a comprehensive study focused towards the potential use cases, architectures, frameworks and promising directions for IBN management and orchestration.

Recent surveys have provided useful insight into the intent-driven network management paradigm, however, they do not cover at least one of the following aspects: (1) intent processing and lifecycle description with a focus towards mobile communications services and networks, (2) intent based network orchestration and management framework for future communications services and networks, (3) detailed analysis of structured intent use cases and domains within the communication networks, and (4) converged architecture for intent-driven networks and service orchestration for mobile communication networks.

### 1.2. Scope and search protocol

This is the first reported structured literature review for intent based networking that gives the reader a comprehensive state-of-the-art and updated solutions related to intent based networking and addresses the concerns with the available surveys [10–12]. A set of research questions are formulated in the following way:

- How can a declarative intent dictate the design, operation and maintenance, as well as the management of networks and the provided services, beyond Fifth Generation (5G) communication networks?
- What are the requirements and role of different stakeholders in the envisioned intent-based networking paradigm? How do these players motivate the utilization of an abstracted policy- as opposed to deterministic policy-based paradigm?
- What is IBN lifecycle in terms of service and network design and which core components are necessary to convert intents into deterministic policies for the network devices?
- What are the key enabling technologies and tradeoffs for the IBN management along with the associated challenges and promising avenues of application?

The objective of this structured literature review is to answer those research questions which also formulate the inclusion criteria for the selection of relevant studies.

Consequently, we formulated a list of relevant keywords as shown in Table 2 encompassing the scope of the design research questions. It generated a total set of 5484 related works published between 2015 and 2021 from Scopus and IEEE Xplore as primary as well as ACM Digital Library, ScienceDirect and WileyOnline as secondary databases. A complete division of the retrieved results from individual databases is shown in Table 3. The duplicates removal, scope filtering, and articles fulfilling majority of the mentioned criterion yielded 60 unique studies contributing effectively to the concept of IBN with different approaches.

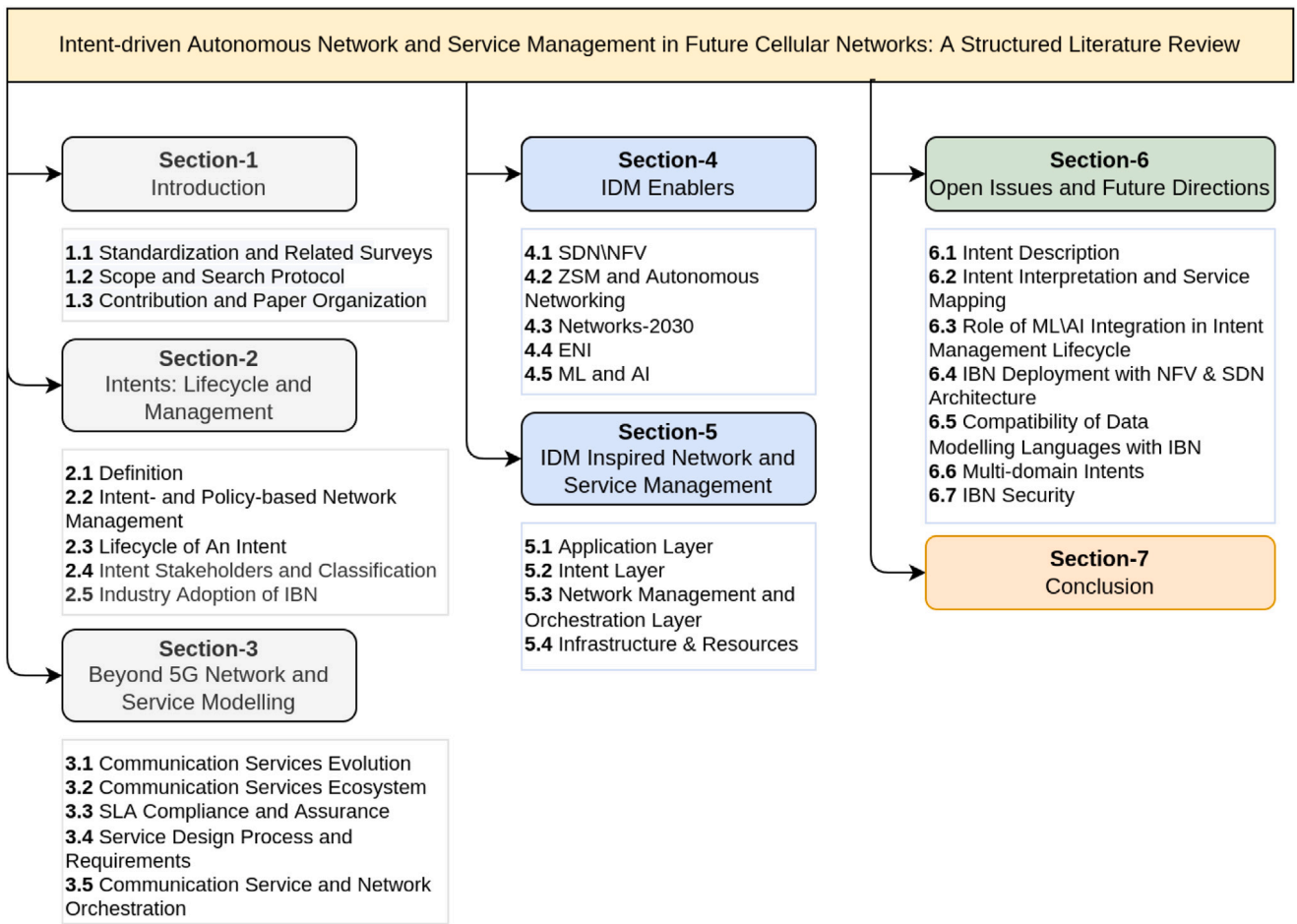

Fig. 1. Organization of the paper.

Table 3
A summary of information sources and search outcome.

| Database | URL | Number of retrieved papers |
|---|---|---|
| Scopus | scopus.com | 3211 |
| IEEE Xplore | ieeexplore.ieee.org | 1158 |
| ACM digital library | dl.acm.org | 56 |
| ScienceDirect | sciencedirect.com | 985 |
| WileyOnline | onlinelibrary.wiley.com | 74 |
| **Total** | | **5484** |

*1.3. Contributions and paper organization*

This study aims at providing a state-of-the-art comprehensive overview of intent based management technologies for the 5G and beyond networks. Prominent contributions are listed as follows:

- Intent semantics including a definition and a detailed description of lifecycle, translation and processing of intents from different stakeholders;
- A discussion of the possible integration (and evolution) of traditional and intent-based network and service management;
- A description of various enabling technologies for implementing IBN management of communication service and networks in 5G and beyond networks;
- A proposal of a converged network and service management framework for the future communication networks with a focus towards SLA and customer Key Performance Indicators (KPIs) along with potential tradeoffs between different intent users and consumers;
- Several challenges and open research directions in order to project possible paths for further studies and a general understanding of IBN management.

For a better understanding of the structure and organization of this paper, we refer the reader to Fig. 1. Section 2 introduces a formal definition and lifecycle description of intents driven management paradigm as well as the difference with traditional policy-based management. Section 3 explains the major players in the 5G ecosystem along with the role of each stakeholder. Service design, orchestration and network orchestration are introduced along with the potential interfacing options for intents from different stakeholders. Section 4 discusses various enablers including ZSM, ENI and AI amongst others to provide necessary features for intent processing and dissemination throughout the network. Section 5 provides a converged architectural framework for enabling intent management of 5G and future networks, with a focus on highlighting the need for SLA and KPI management of service providers and service customers. Section 6 highlights several challenges faced by the intent-driven management of mobile communication networks and potential recommendations and future prospects for alleviating these challenges. Lastly, Section 7 concludes the paper.

## 2. Intents: Lifecycle and management

*2.1. Definition*

An *intent* has been defined as "*a set of operational goals that a network should meet and outcomes that a network is supposed to deliver, defined in a declarative manner without specifying how to achieve or implement them*" [13]. Semantically, an intent can be defined as a specialized type of policy involving several inputs from different parties. This high-level policy can help to manage the network in an autonomous manner without having specific information about the configuration or operation of the network. For systems that are not fully autonomous, intent is typically provided to the network through a central entity involving human intervention. An intent from the service user and the network infrastructure provider sets up a network management perspective that does not require resource and configuration exposure [18]. Moreover, such a system also requires a lower level of control for the management of network functions and service modeling in comparison to imperative policies requiring implementation level knowledge [19]. This distinction allows the networks to abstract the management of network and service level functions without specific instructions requiring device level parameters.

Intents provide a level of abstraction for managing network infrastructures in a domain-agnostic manner. The resulting networks are more flexible and prone to adapting in the face of new service use-cases based on evolving verticals. Verticals can specify the requirements to the network (e.g. need for a particular service with associated quality) to request service availability for a complex service use-case like industrial automation with high reliability and low-latency [20]. The intent translation function maps the contextual information from the incoming user intents to the required services. The information from the intents serves as the input from the respective vertical along with agreed SLAs from the infrastructure provider to requisite appropriate resources and network functionality.

*2.2. Intent- and policy-based network management*

Policy-based network management [21] separates the behavior of different functions from the combined functional requirements of the system. The concept of a policy forms the basis of this design and can be defined as "*a set of rules controlling the choices in behavior of a system*" [22]. The policy is considered as a set of actions triggered according to a defined set of rules when an event occurs in a managed environment. Policies specify the appropriate action to according to various constraints for the network nodes, making the process a set of different control cycles as per defined rules. A central entity provides the policy to be propagated within the network to decide and enforce actions according to the network environment.

In comparison to imperative policy, an intent-based policy is a set of goals in the form of context, capabilities and constraints to be met during network operation in order to satisfy the collective performance targets. The difference between an intent and policy-based management arises from the level of detail specified with the defined goals. Intent-based management consists of providing direction only and does not include additional information to devices regarding the implementation of the rules. The network is not treated as discrete set of nodes, rather as a single entity and the information is disseminated to each device. The relevant information related to the configuration or initialization of physical or virtual functions is processed by network nodes and acted upon if required. Hence, the need for a centralized control is also alleviated, providing flexible and scalable management of network functions. A human intervention is usually required to declare these intents and then the processing is done at device level, regarding the conformance of specific decisions [4].

The key factors aiding the choice of an intent-based management paradigm over a policy-based design are three fold. Firstly, the level of detail and nature of the advised actions/guidelines is critical in choosing the appropriate policy design. Secondly, the implementation scope of the intent-based policy is to be considered as per functional, scalability and lifecycle requirements of network and service nodes. Finally, the notion of decentralized device-level processing of intents, as opposed to the need for a centralized policy-definition and propagation entity.

*2.3. Lifecycle of an intent*

The ability to efficiently manage networks by encompassing user and service provider intents is a complicated process that requires well-defined design goals and models. The intent in itself is an ambiguous entity but when associated with service provisioning or a delivery objective, it becomes critical to a management framework [18]. The specification by IETF [18] makes a distinction for different types of intents according to their lifecycle, namely, transient and persistent. *Transient intents* do not have an associated lifecycle and remain active until the end of the required operation. In comparison, *persistent intents* follow a more structured flow, remaining active unless decommissioned by the intent creator or the network itself. Persistent intents are the focus of this study since they offer scalability and flexibility in intent management [18].

The operational flow of an intent follows its (1) generation in the user domain, (2) processing by different elements to a form acceptable by the network, and (3) deployment in the implementation domain bounded through service orchestration, quality assurance, and feedback mechanisms. This flow consists of two key components responsible for the fulfillment and assurance of the provided objectives of user intents. A general model for the interpretation and dissemination of intents is shown in Fig. 2 that consists of the following domains:

- *User domain*;
- *Processing domain*;
- *Implementation domain*.

*2.3.1. User domain*

The user domain consists of the provisioning of different intents from the users. Several different modules coordinate to perform the recognition of the intent of the users. This is achieved through intricate interactions with different users including the service subscribers, providers and involved 3rd parties. A key goal in this domain is the conversion of intents into a form that is acceptable and understandable by the intent processing domain. The intent description is done with a simple set of instructions by the user through a typical human-machine interface. The intent recognition engine analyzes and interacts with the users to clarify any existing conflicts as well as to complete the modular requirements for the description of a network intent. An actionable user intent is generated and processed with the context of the intent for detailed analysis of its requirements and compatibility for deployment.

*2.3.2. Processing domain*

Intent processing is a vital component of its lifecycle for understanding the needs of the users and providing relevant context to the underlying network functions and resources. The intents are translated into a set of actions, recommendations and guidelines depending on the needs of the user. This intermediary understanding of user intents helps to generate the configurations for different components and functional nodes in the network. Conflict resolution is a key component during the translation of user intents in order to alleviate resource contention during deployment and maximize the feasibility of generated configurations. An optimization framework is required in order to adequately translate user intents into feasible actions. A closed-loop feedback mechanism ensures the compliance of generated intents to the respective SLAs and interpretation of events requiring changes in the intent processing framework. This also provides the flexibility in the lifecycle to modify intent behavior in order to accommodate the dynamic nature of the network and service orchestration processes. The configurations generated for different network domains are distributed after appropriate division of domain specific goals to respective network functions.

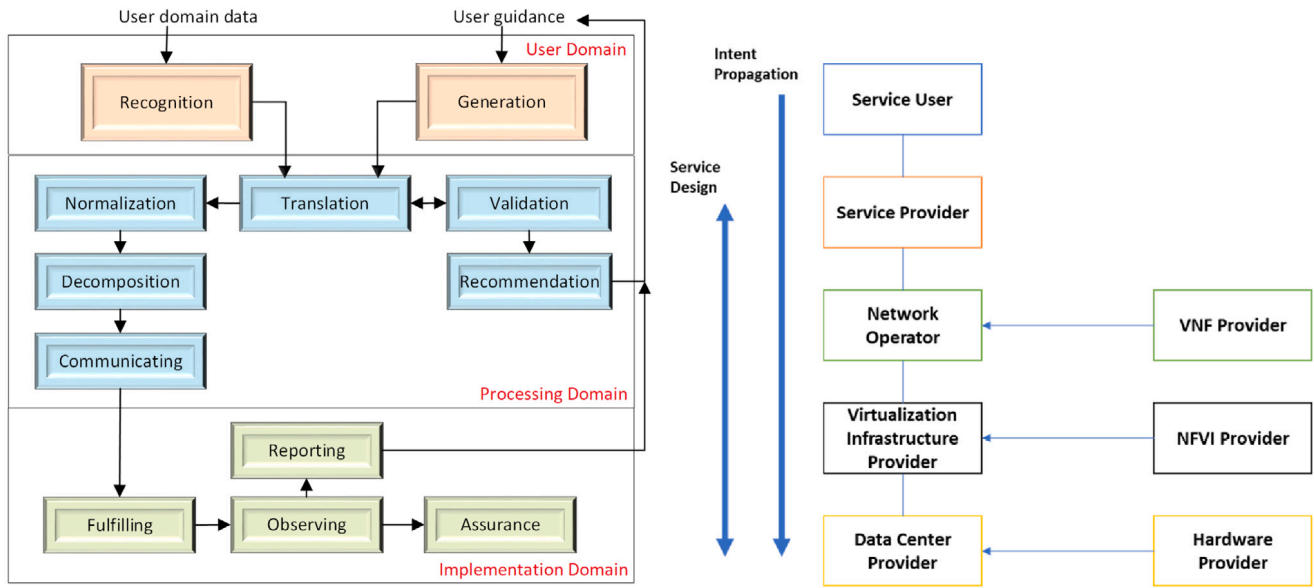

Fig. 2. Intent lifecycle and stakeholders [19].

Intent translation is the conversion of a human readable description of intents, to a machine readable context aided by various Natural Language Processing (NLP) technologies. In a related work, Elkhatib et al. [23] utilize an ontology-based language-mapping method for creating tuples consisting of the objects and subjects of intents. A mapping between user intents and network policies is provided to understand context and deploy user intents. Furthermore, Han et al. [24] proposed a multi-layered intent translation mechanism for virtualized networks with SDN using NLP. The SDN controller and intent engines are combined to provide the required lower level configurations using Representational State Transfer (REST) APIs.

NLP based techniques are complex for the IBN model, where only a limited set of vocabulary is required to represent the user service requirements. Hence, controlled NLP description model is utilized by Scheid et al. [25] to represent and translate user intents with limited number of high-level commands. Then, the refinement process for intents provisions lower level configurations to administer the required service chains in NFV networks. It uses graph based dependencies and clustering to link Virtualized Network Functionss (VNFs) together. Learning based methods are also proposed in the papers [26–28] with NLP to translate user intents using sequential learning, linear regression and label trees to provide context to the user intents. It is however clear that a limited vocabulary can provide the required level of intent context in a scalable manner.

### 2.3.3. Implementation domain

The implementation phase of the intent lifecycle involves several key roles achieved through fulfilling the requirements of the received configurations and directives considering different types of available resources. A key service quality assurance mechanism is embedded in the implementation domain for different types of intents. This helps in assuring the delivery of expected level of performance with well defined objectives and translated directives alongside the interpreted actions from the intent processing phase. An information and feedback relay channel is provided to the intent processing cycle in order to model the incoming configurations, as per requirements of the service model, and bounded by the available resources in different network domains. Intent compliance is also verified during the configuration implementation stage for various involved network functions and the results are made available in the user domain. A key component in this regard is the reporting of deployment outcome to the intent processing phase for closed loop functionality.

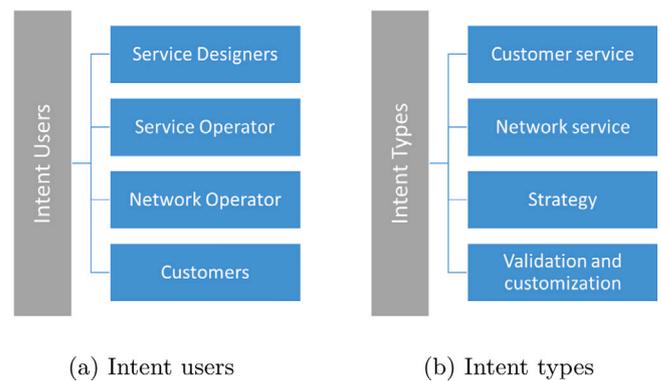

(a) Intent users  (b) Intent types

Fig. 3. Intent classification.

### 2.4. Intent stakeholders and classification

The intents of different stakeholders form a critical part in the intent-based management model serving as the point of conformance for the entire set of network functions. In this context, a classification of the intents and its stakeholders is beneficial and depicted in Fig. 3(a). The players involved in defining and propagating different intents are broadly categorized as the service operators/designers, network operators and customers. The players include their respective objectives during the intent design process in the user domain for generating intents to serve their individual performance goals. In this regard, the customers are more concerned about the provision of multiple service objectives that may be conflicting (e.g. low latency and high throughput). The challenge for the service and network operators is evaluating the tradeoffs and utilizing available slicing and edge computing technologies to alleviate them. The user domain must have the ability to recognize these diverse requirements and generate appropriate intents to orchestrate the underlying network and service infrastructure.

Intent classification [18] is another key consideration to differentiate the scope of the different intents. Fig. 3(b) depicts different intent types based on their functionality and objectives. Customer service intents formulate their objectives according to the requirements of customers in the form of distinct service requests mapped using SLAs during the intent translation. Network service intents provide guidance for the desired configuration, deployment and validation of network functionalities. Strategy intents dictate the implementation of several policies assuring Quality of Service (QoS), resource management, fault detection and resolution as well as the design of different network functions and services. Validation and customization intents offer verification guidelines for other intents in the system and offer additional information for potential modifications of intent handling and conflict resolution in the network.

The segregation of intents based on users and scope provides the basis for defining different intents for ensuring various functions within the intent-based management design [13]. This also defines the roles of different players in 5G and beyond networks to impact the network functionality and orchestration by including their valuable feedback, through intents with minimal knowledge, about the availability of different resources and infrastructures in the lower layers.

*2.5. Industry adoption of IBN*

The adoption of IBN can be powered by the active involvement of networking stakeholders and development of focused solutions with sufficient use cases. We have selected and analyzed the solutions from Huawei, Cisco and Juniper available in different forms in their products. However, a converged solution for multi-domain implementation of IBN is not available.

The first initiative from Cisco [29] has been developing an IBN framework since 2016 as part of its digital network architecture to support multi-domain services and architectures. The primary objective of the project is towards the development and operation of autonomous networks built on SDN with distinctly defined intent interfaces and processing lifecycle. The implementation by Cisco provides a focused approach towards services and applications in data center environments with a focus towards automated analytics and data collection.

Huawei [30] has also been actively involved in the IBN developments with its own flavor of intent-driven networks supporting a dedicated intent engine for the closed-loop control of intent lifecycle. It consists of intent, intelligence, automation, and analytics modules to support the design, translation, deployment and assurance of intents across the network. The focus is towards the implementation in a campus environment similar to Cisco's targeted domain and building upon the SDN architecture. It also provides an Intent API for interaction between the network and intent domain through NETCONF/YANG data models.

Another interesting IBN initiative is Apstra by Juniper [31] to provide a software-based solution with multi-vendor and domain support in a closed-loop automated model. It provides interesting perspective into the composition, reliability, extensibility and scalability aspects of deployed IBN architecture. Another focus is towards the susceptibility to change in the network and ways of dealing with reconfigurations and adaptation with intents as the primary mode on communication between network stakeholders. A context model is utilized for inference and collection of telemetry information from various sources in the network and service domain. This context model is critical in the execution of a closed-loop control and operation of IBN management framework.

**Summary:** The definition of intent is contextual in the sense that it can be stated and eventually interpreted in multiple ways depending on the physical resource availability and network infrastructure. This section summarizes the key concepts and foundations for definition, translation, validation and deployment of intents in a domain agnostic manner. Moreover, different stakeholders and classifications of intents are discussed with a reference towards service and network management use case. A brief view into the available IBN inspired industry implementations provide the necessary insight into the state-of-the-art products for prospective deployments. In the succeeding section, a specific service design perspective is presented in order to enable the creation of context and knowledge base for the translation and verification of intents.

## 3. Beyond 5G network and service modeling

The commercial launch of 5G has rapidly advanced the assessment of its capabilities in utilizing the offered service centric architecture by different stakeholders. This is partially motivated by the projected penetration of the newest flavor of mobile networks capturing around 20% of the market by 2025 [32]. It is also expected that the revenues and investments from the network operators may exceed 1 trillion Euros in Europe alone [32]. Hence, a sustainable development objective for networks beyond deployed 5G benefits not only the service subscribers but also the network and service operators with shared and conflicting objectives.

This section provides an overview of deployed 5G networks and evolving vertical industries in future networks. A communication service design perspective is also considered for understanding the design and optimization of communications services in a Service-Based Architecture (SBA). Moreover, a contextual analysis is provided to enable intents as the primary driver for network and service management.

*3.1. Communication services evolution*

The development of cellular communication networks has moved on with the deployment of 5G networks in the recent years [33]. There has been a shift towards specialized service models and architectures for supporting and disseminating communications services towards various new use cases. The services offered in 5G have been grouped into three main categories: enhanced mobile broadband (eMBB), ultra-reliable and low latency (uRLLC) and massive machine-type communications (mMTC), with diverse requirements in terms of throughput, latency, reliability, availability, and positioning [34]. This has been achieved through various new techniques and methods in the 5G network architecture such as virtualization [35], flexible and scalable management as well as appropriate slicing [36,37] of available network resources.

The challenges associated with the deployment and service expectations related to time sensitive communication and ultra-high reliability have largely been not addressed in 5G deployments. The emergence of AI and Machine Learning as a Service (MLaaS) has enabled complex use cases [38] with learning functions as core components. This helped the evolution of new fundamental communication applications such as Holographic Type Communications (HTC) [39], Multi-Sense Networks (MSN) [40], Time-Engineered Communications (TEC) [38], and Public-Safety Communications (PSC) [40]. These applications will be augmented with new verticals involving compound services that are a combination of multiple fundamental communications services. The complexity of the service environment and infrastructure will certainly require new management frameworks that are able to sense and provide on-demand QoS guarantees to a diverse set of verticals.

*3.2. Communication services ecosystem*

In order to accomplish the diverse service objectives, several trade-offs are to be considered from a communication theoretic and management perspective. The service availability goals for different verticals highlight the importance of flexibility and scalability for the service and resource provisioning mechanisms. The intent based management allows the vertical service providers and customers to aid the provisioning of different services using advanced technologies like network slicing [41] and capability exposure [42] of network to all stakeholders.

The industry and research communities hold a key role in the development of enabling technologies and methodologies for achieving the visualized objectives of different 5G use cases. They encompass different connectivity solution providers such as the network operators, small cell operators, cloud providers, application providers, Over-the-top (OTT) providers. These providers utilize technologies like NFV, SDN, Network Slice as a Service (NSaaS) and Multi-access Edge Computing (MEC) to enable beyond 5G networks for different use scenarios through optimized resource utilization and flexible management. Other key providers are the developers and manufacturers of different components and devices for providing the virtualisation and resource provisioning advantages to the connectivity providers. These technology providers are key to utilizing the emerging techniques and methodologies for the benefit of evolving communication services and networks.

Intent-based management [19,43] is a potential direction towards scalable and flexible management of communication networks by involving users and key stakeholders in the decision making and service provisioning process of the communication networks. Several open-source organizations have a critical role to play in the evolution of the network and service orchestration environment given the premise of network virtualization, softwarization and slicing of network infrastructures. Open Network Automation Platform (ONAP) [44], Open Networking Foundation (ONF) [45], Open Network Operating System (ONOS) [46], Open Daylight (ODL) [47], OpenStack [48], Open-vSwitch [49], Open Source MANO (OSM) [50], and OpenNFV [51] provide the flexibility to utilize technologies like runtime orchestration of network functions, flexible and scalable provisioning of network and service resources for different services to fulfill the requirement of various emerging verticals [52].

*3.3. SLA compliance and assurance*

Provisioning of diverse services is a challenging task due to various factors, ranging from service design to the assurance of appropriate service KPIs. Service Provider (SP) need to ensure performance guarantees to the customers of various verticals and use cases with demanding metrics in terms of ultra-low latency, very high reliability, high spectral and energy efficiency, along with security, privacy, and trustworthiness. These performance metrics lead to the articulation of Service Level Objectives (SLOs) that must be met. Service subscribers and providers formulate the SLA to ensure the conformance of the pre-defined SLOs. The ability to support multiple services requires the network orchestration and management to be flexible and independent of the underlying physical infrastructure [35]. The concepts of SDN and NFV [35] hold the key for implementing adaptable SLAs across the network service provider domain encompassing key network functionalities for different service classes and use cases.

The network and service management approach for the network operators emphasizes the SPs' and users' ability to project their respective capability requirements into the network architecture. These requirements enable the service providers, content providers, and customers to agree upon standard SLAs for different services comprising traditional broadband connectivity and emerging use cases like high precision communications services [40].

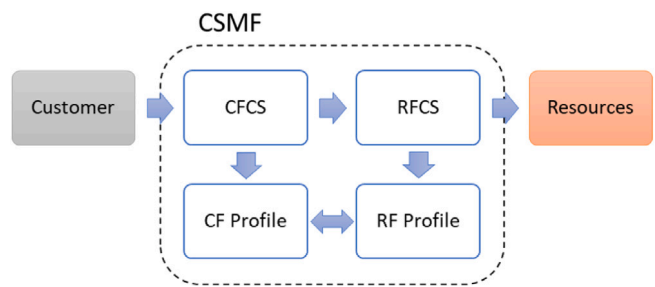

**Fig. 4.** Service design and management framework.

The SPs' and network operators' ability to provide consistent service guarantees and adherence to SLAs is a crucial requirement for a sustained development and performance assurance of service traffic. However, a fundamental tradeoff exists between over and under-provisioning of resources. It requires considerable modifications to alleviate and enable differentiated SLAs in an e2e SBA. To lessen this fundamental tradeoff, network operators and SPs intend to minimize the cost of providing required resources and avoid over provisioning for different service customers as per the defined QoS metrics. However, the heterogeneous nature of the QoS needs for various services requires a context-aware, autonomous network and QoS management design to provide necessary resources to customers, irrespective of their service class and without any resource wastage.

*SLA decomposition* is the allocation of SLA targets for various services or use cases based on network architecture, domains, or users in the form of intents. The advantage of associated specific SLA targets and parameters to different segments provides the much-desired flexibility in resource provisioning for segmented SLAs, in contrast to an enforced uniform e2e SLA. One more aspect of the SLA assurance and decomposition framework is the need for a dynamic and autonomic design to segment SLAs according to complex service requirements projected for the next-generation networks. Intents from the service users, providers and network operators enable the involved decision making required to segment the respective SLA and service objectives in the form of configurable and flexible domain objectives for different network functions. This leads to an accurate provisioning of resources to accomplish the defined service quality and assurance targets.

The mapping of customer requirements from the complex SLAs for heterogeneous services is critical in realizing an effective SLA distribution design across different network segments and customers. A comprehensive network and service management perspective needs an efficient realization of SLAs for the diverse and often conflicting customer requirements in terms of latency, reliability, precision, and connectivity. The need for complex mapping of customer requirements to be automated, forms the basis for a context-aware, automatic, and self-healing SLA management framework. The SLA management framework's automation derives from combinatorial interactions of user intents/requirements, network resource utilization, service structuring, and SLA decomposition analysis.

*3.4. Service design process and requirements*

The concept of a Communication Service (CS) requires a framework to map Communication Service Customer (CSC) and Communication Service Provider (CSP) expectations into a set of ordered decisions for provisioning of demanded resources and network functions as shown in Fig. 4. The communication services should be perceived from the view of CSCs requesting the service and the service CSP providing the required level of service assurance. The Communication Service Management Function (CSMF) provides the required features for provisioning of different customer service instances [53]. This function provides capabilities for processing the CSC's and the CSP's expectations during the service design process leading to appropriate resource allocation. The CSMF further provides the Customer-Facing Communication Service (CFCS) and the Resource-Facing Communication Service (RFCS) for interfacing between the service CSC and the network resource domain.

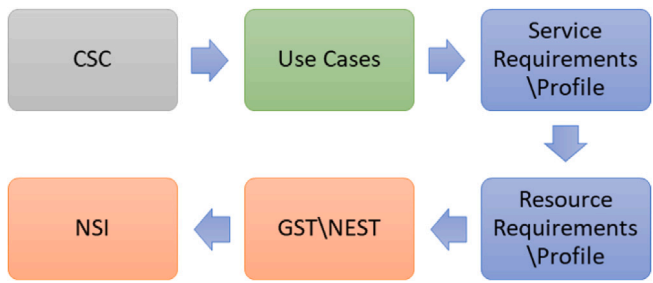

**Fig. 5.** Service orchestration.

The CFCS concerns itself with service design and cataloguing, SLA interpretation and exposure of management capability to CSC. In contrast, the RFCS is concerned with the Lifecycle Management (LCM) of Communication Service Instances (CSI), resource and fault supervisioning and performance assurance for the CSIs. The service design consists of a specification of parameters (type, value range, name) constituting a service profile. These service profiles are created according to the services supported by network and CSPs, later utilized in the lifecycle and management of CSI. The lifecycle of a CSI consists of 4 phases starting from the (1) preparation, followed by (2) commissioning, (3) operation and, (4) decommissioning. Different CSIs are managed by the RFCS along with the respective mapped resources available in the network.

The management of communication services is divided into business, service and network management domains [3]. The CSCs interact and provide their requirements to the CSPs in the form of agreed upon Service Level Specification (SLS) in the *business management layer*. A service request is generated towards the CFCS in the *service management layer* containing information about the required CS for the CSC. The request consists of service specific and customer specific information for the creation of a resource request from the RFCS.

An important concept to consider is the role of customer- and resource-facing service profiles generated during the service management layer. The customer facing profile is constructed from the information provided by the customers in consultation with the CSP and network providers. Similarly, a resource facing profile ensures that appropriate resources can be made available for the service required by the customer. The generated resource requests are forwarded to the Network Management Function (NMF) in the network management layer for the provisioning of required resources or network slices in the case of a sliced environment. The network management layer provisions a new slice (or utilizes an existing slice) to accommodate the requested service for the customer.

*3.5. Communication service and network orchestration*

Various organizations have provided frameworks for orchestration and implementation of resource allocation methods to accomplish the communication service requirements in 5G networks. GSM Association (GSMA) utilizes the concept of Generic Slicing Template (GST) [54] to fulfill the communication service requirements as defined by the RFCS in coordination with the CFCS in the CSMF. Network slicing is utilized to orchestrate resources by the RFCS considering the resource facing service profile for the services requested by the customer. GST is network deployment agnostic and can be utilized in diverse environments. Several Network Slicing Instances (NSIs) [55] can be generated to fulfill service requirements from a single customer, depending upon the propagated service profiles. GST defines various performance, functionality, characteristics, capability and management related attributes. A network slice provider accepts the Network Slice Type (NEST) or GST and triggers the network slice instantiation process in order to perform necessary resource allocations from the network management layer, as depicted in Fig. 5.

An important consideration for the intent based design is the impact of user intents and SLA provisioning on the CSMF functionality in the management domain of mobile networks. Moreover, the nature of requirements passed on from the business layer as well as the manner of enforcement differ for the intent-driven management design. Moreover, the capability exposure in the northbound and southbound interfaces in the network are enhanced to provide additional information to the business layer users as well as the management functions in the network.

**Summary:** It is expected that the fundamental building blocks of future cellular networks would be based on the SBA with an increased focus towards flexible design and deployment of different services. Intent representation and translation require the organization of knowledge related to SLAs, available services and resources. This can only be achieved through active synchronization of information from available sources. In contrast to traditional micro-managed networks, an abstract intent-based approach requires the utilization of novel and emerging methods. These enablers of IBN have been described in Section 4 in order to provide the required context towards a converged intent-driven management model for future cellular networks and services.

**4. IDM enablers and motivation**

Network softwarization paves the way for a high level abstraction of service and network management functions in mobile networks. SDN and NFV enable the flexibility, scalability and adaptability with minimum down time for the network services and nodes. Service design process and user interactions with the network through the NBI dictate the need for an improved management paradigm. IBN management of the network infrastructure provides the necessary benefits in terms of declarative domain agnostic policies. Several enabling technologies have been identified and discussed in this section for expanding the scope of intent-based networking. The discussed enablers include SDN/NFV, ZSM, Autonomous networking, Networks-2030, ENI utilizing ML/AI for network management and operations.

*4.1. SDN/NFV*

Virtualized networks are achieved through softwarized implementations of the network infrastructure and elements through SDN and NFV. SDN provides a programmable network environment to enable a centralized control plane and a distributed forwarding data plane for devices with Openflow [56] as the primary protocol. A SDN controller offers a flexible architecture to configure network wide functionality for enforcing different policies and configurations. Moreover, the benefits of improved resource utilization efficiency, ease of slicing management, security and vendor agnostic design is achieved.

NFV is also a key enabler for intent-based automated network management owing to its ability to abstract different network functions on a resource and functional level. The network elements are termed as VNFs and created on demand according to service and operator requirements. Physical resource utilization is efficient and cost effective for the stakeholders. Moreover, scalability and adaptability concerns are also alleviated with a virtualized infrastructure.

The intent lifecycle consists of description, interpretation, configuration generation and deployment of user provided intents. Virtualization impacts the deployment of generated configurations for different services for any user intent. The service chains and network elements are easier to instantiate and manage allowing enough flexibility and options for swift reconfiguration as per network and resource availability. Moreover, traffic engineering through steering and slicing techniques can be easily implemented in comparison to physical network elements [57].

### 4.2. ZSM and autonomous networking

ZSM provides comprehensive framework for services related to e2e orchestration, intelligence, analytics and data collection in a network agnostic manner [58]. The domain specific counterparts for the management functions within each tenant's scope. Moreover, ITU has recently been working on autonomous networking with a focus towards a ML based service for self-organization of networks [59]. These initiatives can help provide the required automation in the intent-based network management paradigm.

Means of automation must be identified within the intent-based framework in order to enable the intelligent management of network functions and infrastructure. ETSI has outlined a comprehensive framework for ZSM [4] with varying levels of human and autonomous control through different offered service models. ZSM enables a closed management loop [58] for the intent-based networks with deployment, monitoring and assurance of implemented actions. Root-cause analysis of network faults help in identifying the need for reconfiguration of different network functions under varying levels of automation defined in ZSM. Moreover, different automation problems exist within the intent network domain involving the interpretation, description, rendering and assurance of intents for different controllers based on network deployment. ZSM offers to simplify the automation of service models, workflows, configuration policies, and vendor specific needs in the network.

### 4.3. Networks-2030

Communication services have evolved from traditional circuit-switched model to the packet-switched architecture through communication networks. A SBA forms the basis of the prevalent service framework. SBA allows the necessary modular design coupled with the software based networks to allow the introduction of new use cases, verticals and services. ITU recommends a new set of services under the Networks-2030 focus group [40] inspired by the emerging challenges and potential new use cases.

IBN becomes relevant with the influx of new coordinated services, qualitative, tactile communications and in-time and on-time service guarantees. Several gaps have also been identified [60] that can be fulfilled by utilizing an intent-based framework for future networks. SLO-aware network service interfaces and negotiation allow the users to specify the SLO when negotiating with the network for service compliance. This can be ensured by IBN through user and network involvement in the form of SLAs. Coordinated services are supported where multiple services requested by users are dependent on each other. This is ensured through the negotiation of requirements and closed loop feedback from the network to the intent users. Moreover, the gaps related to programmability, lifecycle agility, customization and assurance of emerging services can be resolved with IBN coupled with SDN/NFV frameworks.

### 4.4. ENI

The dynamic nature of modern networks requires the adaptation of services, network infrastructure, configuration and management policies in a closed loop. ETSI ENI helps alleviate the challenges related to the automation of complex decision-making, optimized service LCM, and network state. IBN shares the common goal of user-driven intelligent service, configuration and network management with ENI [8]. The ENI provides an architectural framework to implement functions for context awareness, cognition management, situational awareness, policy management, model-driven approaches, and knowledge management to provision the experiential intelligence. Moreover, ENI defines different levels of autonomy for networks with Cat-4 and Cat-5 (Table 1 in [7]) specifically with intents as human-machine interface. These networks with partial and full autonomicity have deep awareness of network status and are capable of performing self-initiated decision-making and optimization during operations and maintenance phases. In addition, ETSI [8] proposes an intent policy management framework as well within the ENI domain in order to extract the imperative policy from the declarative intents. Utilizing the described ENI functions, an improved interpretation of different types of intents can be provided to the cognitive management function of ENI.

It also proposes different architecture frameworks [9] for utilizing the defined levels of automation with AI functional blocks. This flexibility allows varying levels of human-machine control of the network management according to different use cases as well as compatibility amongst them. ENI can form the basis of the complete LCM of the network management, from intent processing to the resource modeling and service deployment.

### 4.5. ML and AI

Inference forms a critical component in the intent processing and service design lifecycles. It is also a key requirement for the expected level of autonomy in future networks, ranging from human assisted methods to being fully self-sufficient. ML and AI are a driving force in achieving the intent level inference given the contextual awareness and sufficient data availability from various network and intent functional blocks. Following are some potential areas of interest that require AI in the composite IBN framework (Fig. 6):

- Extraction of service primitives from intents,
- Optimization of service orchestration,
- Intent-to-service mappings with historic user behavior,
- Orchestration and configuration profile generation from service descriptors,
- Intent and service assurance with proactive monitoring of resources, user expectations and network environment.

The scope of learning-based automation in IBN framework is not limited to only the mentioned applications above. The needs of modern networks are continuously evolving as highlighted by the recommendations from several Standard Development Organisationss (SDOs) [4,9,59].

Dedicated ML functionality is a key component in any future autonomous networking framework to ensure various levels of autonomy [4,7]. MLaaS [61] has been proposed as one alternative. However, focused efforts are required in order to overcome the associated challenges [62] in implementing such a functional block in modern networks.

**Summary:** In this section, we have reviewed the enabling technologies for realizing an intent-driven service orchestration and network management. It is observed that the creation of context and knowledge base is a significant challenge that can be alleviated using input from different stakeholders including service and network infrastructure providers, network operators and service subscribers. ML provides the means to extract the key components from a declarative intent from the subscriber and performs the mapping between requested and available service options. ZSM, ENI, and SDN provide the necessary possibilities of integrating IBN in standardization efforts for future autonomous networks. By utilizing the aforementioned enablers, we have proposed a detailed description of an IBN infrastructure for future cellular network deployments in Section 5.

## 5. Intent-based network and service management

Network management and service provisioning, with varying levels of SLAs and QoS expectations from various stakeholders and subscribers, can be accomplished in a flexible, dynamic and intelligent manner by using different user intents. The softwarization of network and resource infrastructure through SDN and NFV provides a suitable implementation scope for an adaptive, user centric and high-level management design supported by intent-based control loops. The network and service design infrastructure is influenced by the intents generated from the users, keeping in view the agreed SLAs and service profiles available with the service providers.

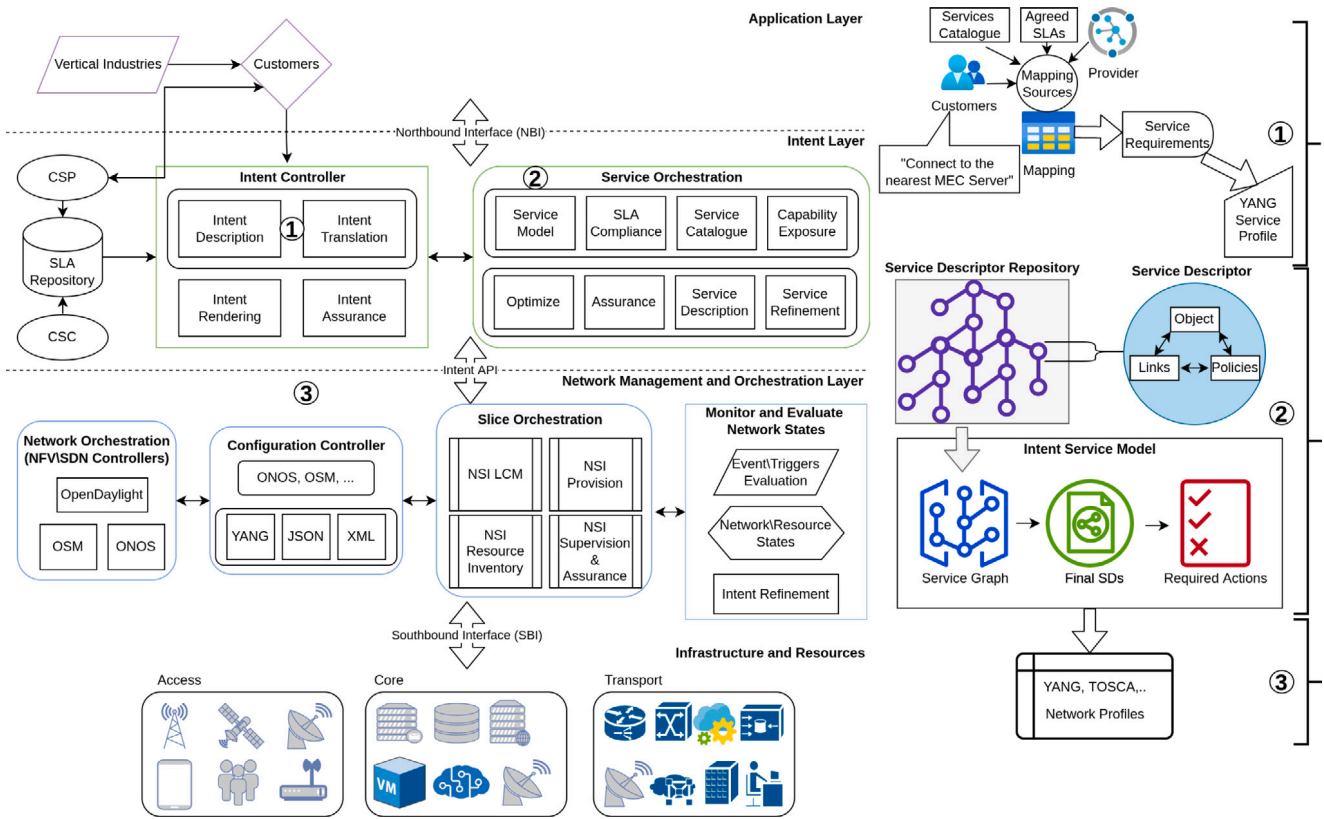

**Fig. 6.** Intent-driven service identification and network management framework.

An intent based architecture for the network and service management is proposed in Fig. 6. It consists of the application, intent, and network management & orchestration layers along with the underlying resources. A brief description for each layer is provided in the following subsections.

*5.1. Application layer*

The application layer consists of the various vertical use cases and their consumers. The consumers are provided with exposure of underlying network infrastructure and services through the Intent NBIs and SLAs for infrastructure and service availability awareness. Intent processing aids in understanding the scope of different domains and provides e2e expectations of services available through service providers.

The representation of intents by users is an important design goal for the IBN framework since it involves a tradeoff between complexity and scalability of the network. The intents can be represented in a machine and human-readable form using ontology-based [63], vocabulary restricted [25,30], or NLP methods [24].

A constrained intent description model is adopted by Elkhatib et al. [23], that provides a layered approach to specify intents using three basic primitives construct, transfer, regulate, for different phases in user-defined intents. Each primitive module then constructs a specific type of intent to perform the required task in the network. On the other hand, Han et al. [24] provide pre-defined intent objects by following the ONOS intent framework. The topology, endpoint and chaining objects are provided in accordance with the requested intents by the users. Moreover, conflict resolution plays its part in order to resolve any disputes during the translation and processing of the intents.

Constrained Natural Language (CNL) restricts the broad domain of NLP into more machine readable scope in order to alleviate the extensive processing, scalability and adaptability requirement of intents. This approach has been explored by Scheid et al. [25] providing an alternative method to refine intents by dividing the specified requirements into essential and operational specifications. A Softgoal Inter-dependency Graph (SIG) is constructed to represent these formulated modules. A quantitative assessment provides the most appropriate SIG for the intent with different contributing requirements to be implemented in the intent layer.

NLP provides a user-friendly and high-level abstracted view of the intents as shown in [26–28]. In the study, Jacobs et al. [26] utilize a chatbot based dialogue framework to interact with the user and construct an intent model using a proposed abstraction language. Key performance goals are extracted and forwarded towards the intent layer for translation using Recurrent Neural Network (RNN). In another work, Chao and Horiuchi [27] describe the intents using a restricted vocabulary model and any vagueness in descriptions is processed using rule-based parsing and matching model. The user definitions are saved and utilized in the intent refinement process for subsequent requests. In a related work, Kang et al. [28] consider label namespaces to maintain relationships and interactions between different network entities eventually aiding intent composition and translation. The labels are arranged as automatically generated label trees representing the complex dependencies between the network nodes and the dynamic nature of the environment. The user can use these labels to define the intents instead of considering low-level network details and relationships. Furthermore, the exposure of network capabilities through the NBI to users provides the necessary context for intent description as shown in [63–66].

## 5.2. Intent layer

The high-level abstraction of network capabilities and available resources is provided through the intent layer. Consumers specify their expectations in the form of intents to the Intent Controller at the start of an intent lifecycle.

### 5.2.1. Intent expression and NBI

The ability of intent ingestion into different network stakeholders depends on the appropriate interfacing in the management and control domain. The intent sources (e.g. service subscribers) follow a declarative NBI model for the design of an intent providing the required interface towards the intent controller and the orchestrator. This interface is independent from the specific implementations of the service provider and subscriber input, thereby it provides a common interaction design for a diverse set of providers. Huawei proposed the NEMO language [30] for describing intents. A key feature of this intent NBI is the utilization of mappings to facilitate arbitration of information between intent sources and intent providers. These mappings enable providers to interpret intents in terms of system-specific model leading to efficient orchestration and deployment of services.

The Intent Controller defines the perspective of the user intents through translation and cross referencing them with the service and operator catalogues with SLAs as well as service and network specifications (Step **1** in Fig. 6). The mappings are established between consumers and network providers in order to help the translation of the intents in the form of lookup tables and catalogues. Human readable consumer terminology is converted into machine readable form for the network and service provisioning. The mappings are dynamically updated according to the changes in the state of the consumer and provider capabilities, including the available resources in the network. Afterwards, intent perspective is utilized for service and network orchestration by the network controllers for example in SDN and NFV functional domains. However, it is domain specific and has limited phrases to represent user intents.

Several different models have been proposed involving the exposure of underlying network resources and functions through a descriptive model to aid in the intent translation process. The study by Singh et al. [67] provides a multi-layered approach to classify different types of traffic and improve data dissemination through the vehicular network environment. The intent processing is done using an ONOS intent controller along with an AI-based optimization engine. A user behavior driven model is adopted by Esposito et al. [66] for intent translation into distinct well-defined network policies. A NBI abstraction layer is proposed with a NLP based Gherkin language for intent definition with certain keywords embedded for intent interpretation and mapping onto low-level configuration policies. Moreover, the study by Chen et al. [64] reports the utilization of a human-readable query language for user intent expression and web ontology and resource description framework to aid the processing of user intents for a VNF chaining application. The role of low-level network capability and resource state exposure to the intent processing layer is validated in this work for providing service composition and deployment. The work of Kiran et al. [63] performs an investigation into the NBI interface providing useful insights into the provisioning of differentiated services. A high-level abstraction intent translation module receives the intents from users and queries the network requirements for different applications using Resource Description Framework (RDF) graph repository.

Intent rendering involves the creation of a service model through the generation of service objects and service descriptions to be forwarded to the configuration controller for deployment. Intent assurance is responsible for monitoring the status of intent compliance as per deployed configurations and assist in initiating reconfiguration on a service and network level, based on observed events in the underlying layers. The focus of the intent layer is to provide meaning to the customer service intents (referred to as user intents). The Intent Controller is also involved in generation of network intents to provide context to the network orchestration and management for deploying the required services from the user intents. This is accomplished through the intent APIs provided between the intent and network layers. The network represents intents in a machine readable form, with Yet Another Next Generation (YANG) as a possible modeling language. Similarly, some strategy intents are also generated to help in defining the behavior of the intents of various types for different functioning elements. Another key intent is the validation and assurance intent that helps in modifying the behavior of user intents due to the feedback received from the intent monitor module.

### 5.2.2. Service model and orchestration

Intent-based network management is built upon Intent APIs to provide service models across different domains, network controllers and in an e2e manner. The service deployment complexity is reduced with the software-based network architecture ensured through NFV and SDN technologies. A modular approach can help resolve the complexity of service design process by providing service agnostic components to compose services. This makes the service modules more reusable and portable to any type of service orchestration domain including intent-based model.

Once a service request is received from the intent controller, a service model is generated using different Service Descriptor (SD) deemed necessary by the intent controller. A service descriptor consists of the service specification object, the links to other SDs and the polices defining its behavior as depicted in Fig. 6. The policies can be defined by relationship to other SDs, through functions and conditions to other SDs, and through explicit declaration by the user intents. The state evolution of the SD states is dictated by a Finite-State-Machine (FSM) with design, reserve, provision and active state values. A set of required actions in the form of service requirements and specifications are generated from the selected SDs. These actions are specified in the form of service description models for the underlying domain specific network orchestrators and controllers as depicted in Fig. 6.

The work by Paganelli et al. [68] explores the Network Service Description (NSD) model for intent-based orchestration of network functions in SDN. A layered service-oriented model is devised with distinguished business and orchestration scopes. Topology and Orchestration Specification for Cloud Applications (TOSCA) [69] and YANG [70] models are used for service description and automation of service lifecycle. Extensions to the ETSI NSD provide the necessary information exchange between the intent and network controllers for service orchestration. A NBI and e2e service model architecture is proposed by Cerroni et al. [71] for the IoT use case. A Proof-Of-Concept (PoC) is also investigated over an OpenFlow based SDN testbed with promising results in terms of response times for the service requests. Service chaining is performed using NFV, Virtualized Infrastructure Manager (VIM) and Management and Orchestration (MANO) through intent-based abstracted service requests.

In a related study, Rafiq et al. [72] investigate a platform agnostic service design with implementation for OSM and M-CORD orchestration frameworks. Service orchestration and VNF modeling are performed using an IBN manager and policy configurators for different platforms. Users directly design the intent contracts using a GUI and the intent manager interprets and deploys them using a vocabulary store with historic intent contract data. TOSCA and Network Slicing Template (NST) [54] models are used for M-CORD and OSM orchestrators for deployment of intents. Moreover, the model proposed by Scheid et al. [25] uses SIGs for service requirement mappings and clustering of relevant VNFs in an automated manner from user intents by evaluating the available network functions and requested applications.

*5.2.3. Service assurance and optimization*

Service model consistency is ensured by the intent controller through policy evaluation of the generated service object graphs. The service impact model generated during service orchestration is updated, when changes in the network and resources are reported through the monitoring module in the network management and orchestration layer. The service model is optimized according to the updated states and a new service graph can be generated if required. The closed loop feedback from the monitoring module is critical in providing dynamic service lifecycle for user intents. The change in service state can also be caused by the user based events like intent refinement through a dialogue to clarify any conflicts between different intents or lack of feasibility of the requested intent.

Various frameworks have been proposed for service assurance and conformance with the user intents and requirements. In a related work, Abbas et al. [41] propose a Generative Adversarial Network (GAN) based resource management framework for IBN with network slicing for different services in 5G network. OSM network orchestrator coupled with a FlexRAN controller manage the slices configured through high level intents from the users. GAN provides with the ability to predict future states of the resources in the network and allow refinement of user intents and network orchestration. In a recent study by Szyrkowiec et al. [73], the process of service provisioning is automated with the utilization of REST [74] NBI and action mapping for user intents using the intent compilation module. The actions form the basis for the service model with NBI and SBI interactions using the REST and Network Configuration Protocol (NETCONF) [75], REST Configuration Protocol (RESTCONF) [76], YANG [70] modeling languages. A recent study by Khan et al. [77] proposes a GAN based Service Graph (SG) generation to aid Deep Reinforcement Learning (DRL) based optimal SG selection process based on user intents. A VNF graph is generated and improved by GAN whereas the DRL engine provides its input with the optimal policy for deploying the graphs without any conflicts. A multi-domain orchestration scenario is also provided for deploying the optimized configuration policies.

*5.2.4. Role of intent API*

A major objective of intent-based systems is to infer the intent of the subscriber instead of using a precise description of the objectives and expected deployments. Intent APIs allow the intent providers to understand, correlate and interpret the required services and network nodes in order to accomplish the intent. This can be done in multiple ways depending on the underlying network infrastructure, data description languages and service models. In the proposed framework, the intent API provides two fundamental functionalities in the context of a cellular network environment. First, the exposure of underlying network infrastructure towards the intent controller and service orchestrator is performed using different data models (YANG [70], TOSCA [69]) in the service repository. Second, the intent controller and service orchestrator ensure the conformance with the relevant domain specific models for the creation of service descriptors [13] and network functions in collaboration with the network controller in the underlying network management and orchestration layer.

*5.3. Network management and orchestration layer*

This layer constitutes the network orchestration, configuration controller, slice orchestration and monitor and evaluates network states modules. The network orchestration constitutes different domain controllers and orchestrators, including NFV MANO, SDN, and slicing controllers interfaced to the intent controller through the intent APIs. The network layer expects the description of service orchestration and device level service models to be integrated seamlessly using the varied orchestration and deployment support. The service requirements in the form of TOSCA [69], HOT [78] orchestration models are received by the configuration controller for generating YANG, Extensible Markup Language (XML) [79] or Yaml Ain't Markup Language (YAML) [80] based models for different network devices with exposure to the resource state through OpenAPI [81]. Moreover, the network and intent reconfigurations are also handled by the configuration controller in coordination with the monitoring and evaluation module. A set of network orchestrators (ODL, ONOS, OSM) are utilized to abstract the resource state and deploy the generated configurations for different user intents in the form of network slices, VNF graphs or SFCs.

The study by Abbas et al. [82] utilizes multi-platform orchestration with OSM for core network and FlexRAN for the RAN part of the network. The intent interpretation is propagated to the configuration generator for the respective controllers along with an active monitoring module implemented through GAN. Moreover, the work by Addad et al. [83] focuses on interfacing between intent layer and network management layer and utilizes an ONOS based SDN controller to manage intent types and lifecycle. The orchestration and models have been designed for the cloud applications and traditional network configuration tasks. For the intent-based operation, certain improvements are mandatory in order to enable their utilization. A comparison of orchestration and configuration models suitable for IBN is provided in Table 5 along with highlighting a set of improvements.

The monitor and evaluate network states process is responsible for interacting with the domain controllers and orchestrators for assuring the compliance of propagated service deployment requests and intents on a higher level. It also interacts with the infrastructure regarding the resource status through the SBIs and network orchestrators to get state information necessary for providing the directions to the intent controller for triggering events. Consequently, any required changes are pushed by the intent controller via intent layer modules for appropriate service refinement and optimization. This is one of the avenues for the utilization of ML and AI based methods to monitor, predict and adapt the intent guarantees in order to maintain the requested level of performance for different provisioned services. This is evident from a study by Tsuzaki and Okabe [88], where intent refinement is accomplished in a reactive manner by the network administrator through monitoring of different KPIs in an Internet based environment with NETCONF for device level configurations. A related study by Jacobs et al. [89] focuses on a self-driving use case to model intent translation and update process using a sequential learning model using RNN. Moreover, a dialogue enables the user and intent processing engine to interact when required in order to improve the learning of the deployed ML model.

*5.4. Resources and SBI*

The network infrastructure provides the required physical compute, storage and communication resources in order to provide building blocks for the software based network and service orchestration. The resource information is shared through the SBI to enable resource status monitoring and events management through the network orchestrators and controllers. The SBI interface also provides the required exposure to the monitoring module responsible for the closed loop feedback management for the intent layer. Moreover, the resources are distributed according to the domain specific definitions amongst the access, core and transport parts for different types of offered services in the network through different providers. The resource description is a key consideration keeping in view the diversity of domain controllers and orchestrators. However, a virtualized and softwarized network environment takes precedence as depicted by some resource description proposals [42,95].

**Summary:** We explored a wide range of relevant projects utilizing intents as key sources of policy design for orchestration and management of networked services. The key challenges in a future cellular network driven by intents are the collection and understanding of data from various sources (resources, nodes, services, SLAs). This has been explored in various related studies summarized in the proposed architecture and organized in Table 4. Moreover, the proposed IBN model is extendable to multiple domains by adjusting the domain specific data models and network architecture.

**Table 4**
A description of IBN architecture scope and related works.

| Layers | Functions | Features | References |
|---|---|---|---|
| Application | Intent modeling and NBI | → Description of user intents using ontology, various NLP, and vocabulary methods.<br>→ Customer and business-oriented design of intent representation.<br>→ Benefits in terms of scalability and processing intensity for human and machine-readable forms. | [23–28,30,84] |
| | Compatibility and Capability exposure | → Conformance of user intents with the underlying network and intent infrastructure.<br>→ Exposure of network capability to users in order to provide required information for intent modeling. | [25,63–66] |
| Intent | Intent processing and perspective | → Translation, processing and context generation for intents with exposure to network capabilities.<br>→ SLAs and information from service and network providers is available for the contextual evaluation. | [63,64,66,67,85] |
| | Automation and Refinement | → Update the status of low-level resources and functions for any required changes in the intent context.<br>→ Dynamic behavior to support and process abrupt changes in the environment.<br>→ Conflict resolution for different intents in case of contended resources and demands. | [41,69,73,77,78,86] |
| | Service Model and Orchestration | → Service modeling with service graphs and data modeling languages along with LCM.<br>→ Generation of low-level configurations for the domain specific network orchestration.<br>→ Exposure of low-level resource level information to enable required configuration templates. | [25,64,68,71,72,87] |
| Network management and orchestration | Network Orchestration | → Intent rendering takes place with the allocation and LCM of different network functions.<br>→ Several different domain specific orchestrators can be utilized like OSM, ONOS, OpenDaylight. | [82,83] |
| | Intent Adaptation | → Monitoring functionality for enabling the automation and dynamic orchestration of services via intents.<br>→ Exposure of resource layer through resource description models to enable trigger conditions. | [88,89] |
| | Intent Deployment and Configuration | → Deployment of user intents using domain-specific controllers.<br>→ Resource level information for mapping the VNF and SFC to the appropriate nodes. | [70,72,90,91] |

**Table 5**
A distribution of different available and expected features in Data Models for IBN.

| Model (SDOs) | Features | Role in IBN lifecycle | Required improvements for IBN lifecycle | |
|---|---|---|---|---|
| | | | Service model | Service orchestration |
| TOSCA (OASIS) [69,92,93] | – Declarative & imperative, human-readable & domain agnostic<br>– Service, resource, & inter-dependency modeling<br>– Validation & fine-grained modeling of services<br>– Can be represented in XML, YAML or YANG | – Service orchestration & LCM<br>– YANG model instantiation for service components<br>– Service assurance through adaptable configuration management | – Multi-level graph model (graph, tree-based)<br>– Inheritance for service descriptors (complex service design from basic templates)<br>– Portability & reuseability of basic service descriptors | – Language semantics definable internally<br>– Dependency models based on policies<br>– Dynamic service graph creation |
| YANG (IETF) [70,94] | – Machine-readable & domain agnostic<br>– Device-level configurations & instantiation | – Representation of service requests as dependencies for intents<br>– Service deployment & configurations as per service graph | – Coupling between services and service components<br>– Structured naming convention<br>– Automated deployment & upgrades | – Deployment state model for service descriptors<br>– Dependency analysis of services in runtime<br>– Automated modification of deployed services |
| HOT (Openstack) [78] | – Declarative, human-readable<br>– Service LCM, description & orchestration<br>– Static and limited support for adaptability & new models | – Service orchestration & LCM<br>– Not suitable for complex services & scenarios | – No dependence on underlying protocol stack<br>– Utilization of dynamic service blueprints for service design | – Dynamic LCM of services<br>– Resource models & exposure |

## 6. Open issues and future directions

The conversion of abstract declarative intents into definite configurable policies is one of the biggest challenges and design goals in IBN systems. This is achieved by embedding automation in the network operation and service-design process with the aim of making the network more deterministic by utilizing optimized intent descriptions and service mappings. After carefully reviewing the relevant literary sources, we have identified several open issues that require further investigation and concentrated efforts in order to realize the IBN management framework.

*6.1. Intent description*

The declarative nature of intent descriptions from the service subscribers requires abstracted versions of service requirements in terms of context, capabilities and associated constraints. Machine readable format for intents is preferable for scalable intent processing in complex heterogeneous service environments. However, service subscribers are more comfortable with a more human readable format that can be converted to a machine readable format by leveraging NLP techniques. This dilemma highlights one of the fundamental processing and scalability tradeoffs for the LCM of intents as well as the provisioning of required resources to the requested services. It is expected for the intents to be human-readable at the business and subscriber level and more machine understandable as the intent flows through the management framework towards orchestration. Several approaches have been explored including pure unrestricted language [66], restricted vocabulary [25], ontology [23] descriptions amongst other hybrid methods [26,27]. The intent representation problem still needs attention through contextual NLP methods in order to provide scalable solutions for human-readable intent description models.

*6.2. Intent interpretation and service mapping*

Service and device level interpretation of declarative intents requires the knowledge of network capabilities, service provider offerings and resource level exposure [13]. Service models also need to modify in order to accommodate the IBN requirements in terms of declarative parameters (context, capabilities and constraints). A service design process should be offered in a modular architecture in order to construct flexible services considering the context and resource availability for different intents. The concept of microservices [96] adopted from software engineering can provide the decomposition of the service design lifecycle into individual sub-services in order to construct the services according to context in IBN.

MEF [97] has worked extensively into the realization of an automated Lifecycle Service Orchestration (LSO) and provides a model based intent-driven service design and orchestration framework. The proposed framework utilizes Domain Specific Language (DSL) to translate and compile intents into network and device configurations utilizing the context information from respective domains. The service requirements are extracted in multiple iterations with each pass extracting useful contextual information from the intent description completed upon the construction of the service description graph (tree). The standards have been more focused towards web services and further investigation is required in order to confirm the feasibility for communication services.

*6.3. Role of ML/AI integration in intent management lifecycle*

As described in Section 4, autonomous behavior in network management and service orchestration via intent based policies can be envisioned through different ML and AI techniques. Intent description model, mapping intents to service requests, LCM of intents and services are some of the potential problems that require dynamic mapping mechanisms with learning modules. NLP helps in understanding the intent structure and context through traditional Sequence-to-Sequence (seq2seq) learning models like recurrent neural networks (RNN, Long-Short-Term Memory (LSTM), Gated Recurrent Unit (GRU)) [98] and advanced models including attention [99], Bidirectional Encoder Representations from Transformers (BERT) [100]. The context and declarative parameter retrieval leading to the service extraction can be achieved through a Convolutional Neural Network (CNN) iteratively considering the service domain and environment as well as information related to SLAs and supported services. The introduction of MLaaS [62] notion provides the required learning in different aspects of network operations and management as an on demand feature. Learning as a core functional block has also been included in the ETSI proposed ZSM, ENI model frameworks for autonomous networks of future with differing levels of automation. However, focused studies related to the learning based problems in IBN model are still lacking.

*6.4. IBN deployment with NFV & SDN architecture*

The softwarization of network functionalities and resource management have allowed dynamic orchestration frameworks in the form of NFV and SDN. However, the orchestration and network operations are dictated through definitive imperative polices [19] with pre-defined outcomes lacking truly autonomous behavior. IBN provides the ability to different stakeholders [13] to influence the design, operation and provisioning of network resources to enable virtually any type of service in future networks through abstract policies. The mapping between declarative parameters from different stakeholders and the LCM of VNFs and SFCs as well as the software defined network nodes is a key challenge in softwarized infrastructure. ONAP has explored an infrastructure for integrating intent-driven configuration and orchestration of various services in 5G. Moreover, similar efforts have been reported in MEF PoC [97] for the SD-WAN and Network Slicing (NS) use cases for the 5G services paving the way for the application and deployment of different IBN processes in various PoCs in future studies. However, several efforts are required in different domains from the perspective of different stakeholders, in order to realize the notion of IBN as a key technology for driving domain-agnostic networks.

*6.5. Compatibility of data modeling languages with IBN*

Data modeling languages provide the means of describing the structure of services, orchestration process and the device level requirements for the implementation of a network and service orchestration lifecycle. YANG and TOSCA based models have been utilized to represent the intents, processed service descriptions and orchestration and device level objectives for different services in various studies [9,58,97] involving declarative intent policies. Several improvements have to be embedded in the language models leading to the specification of DSL for different domains as depicted in Table 5. The envisioned changes to the available description models are relevant to the service model and automation in the IBN centric solutions. The service model requires the introduction of multiple levels of service models in order to allow a flexible abstraction model for individual services. The reuseability of service description models is also important in order to accelerate service design process. Moreover, dependencies between service models and components need to be dynamic in order to modify different service instances in runtime. IBN centric structuring of naming and catalog information in TOSCA, YANG data models is also required for utilization in the declarative policy based network management frameworks.

On the other hand, service automation requires the definition of language semantics, dynamic dependency and service graph models, service modification routines, followed by resource and device level exposure for orchestration. Language semantics for automation are required for representing imperative policies driven from intents and for the LCM of services and orchestration yet they are missing in YANG and TOSCA models. Inter-dependencies between service components and constructed service graphs must be designed, created and reflected in the representation modules to propagate the required information during resource provisioning. Service descriptors and instances should be flexible enough to allow runtime modifications in order to adapt and modify existing policies to the changing runtime environment. The resource models described by Wu et al. [42], and Paganelli et al. [68] also need to be modified since they are not sufficient to support the automation in operation and maintenance of networks according to the abstracted policies.

*6.6. Multi-domain intents*

The implementation of a centralized intent-driven orchestrator poses scalability and contextual interpretation issues with intra-domain and inter-domain intents. This is due to the limited understanding of global context and knowledge necessary to build multi-domain translation and mapping models for the local intents [101]. There is also a potential conflict between local intents and inter-domain provider intents due to varied exposure of information between different domains. For instance, the treatment of an application intent in different domains of a cellular network require the knowledge regarding ownership, permissions and available resources in each domain. This is also coupled with the need to distinguish and prioritize the processing and deployment of critical services versus local intents [102].

The process of intent consumption involves the description, representation, translation, validation, deployment, and assurance [19]. These are complex tasks to perform even for a single orchestrator having complete knowledge of the state of the network and services. In a multi-domain environment, an orchestrator possesses only the local snapshot of the global state. The optimization and globalization of intents across multiple domains poses considerable challenges in addition to the ones faced by a centralized intent orchestrator.

*6.7. IBN security*

The analysis of security considerations in an IBN environment is critical due to the enhanced role of intent sources and subscribers compared to a traditional network. One such concern is the assurance of secure provisioning of service flows during the intent translation process with service dependent levels of security. Moreover, the level of influence from the intent users needs to be identified and modeled according to different roles in the network, for example subscriber, provider, operator. The potential of malicious intent requests exists requiring a complex intent verification policy in order to control the level of access and provisioning for different user groups. Another key consideration is the transparency in the ownership of intents within the network domain which can also lead to secure processing and content resolution as well as detection of malicious intents [103].

The implementation of security policies in the network can also be simplified by using intents with minimized risk of configuration mistakes. Expressing security policies for different components and modules in the intent lifecycle provides the basis to utilize the existing processes in the deployment phase [104]. Large-scale network security can also be ensured with intents serving as the key deployment interface for different types of networks and devices [105].

## 7. Conclusion

In this article, we analyzed the state-of-the-art in intent based network management and service orchestration. The potential benefits of the IBN infrastructure are explored in relation to the future networks and evolving service types. The progression and lifecycle of intent are explained in the context of context-aware abstract policies and conversion to low-level imperative policies. The role of abstract-policies in the deployment and management of scalable network functions and services is emphasized along with the need for learning based functionality embedded in the network infrastructure.

This review provides the necessary insight into various investigations employing IBN as the main driver leading to autonomous network operations and management. We included different approaches for orchestration and network management, where, IBN is utilized as the core policy for driving the management of network functions and provisioned services. Detailed analysis of the relevant studies highlight the need for ML/AI as functional elements to provide learning techniques in order to improve autonomous behavior in the control loops. However, there is need for concentrated efforts in order to identify the means and level of automation appropriate for different types of intents and requested services.

A converged network management framework is proposed consisting of the closed control loops for the processing and interpretation of intents, as well as service identification and orchestration. This is complemented with the discussion of different data models for suitability in the orchestration and service modeling lifecycle from originating user intents. Several challenges and open issues as well as future directions are also provided in order to drive the efforts towards IBN management as a viable option for future communication networks.

## CRediT authorship contribution statement

**Kashif Mehmood:** Conceptualization, Data curation, Formal analysis, Funding acquisition, Investigation, Methodology, Project administration, Resources, Software, Supervision, Validation, Visualization, Writing – original draft, Writing – review & editing. **Katina Kralevska:** Conceptualization, Data curation, Formal analysis, Funding acquisition, Investigation, Methodology, Project administration, Resources, Software, Supervision, Validation, Visualization, Writing – original draft, Writing – review & editing. **David Palma:** Conceptualization, Data curation, Formal analysis, Funding acquisition, Investigation, Methodology, Project administration, Resources, Software, Supervision, Validation, Visualization, Writing – original draft, Writing – review & editing.

## Declaration of competing interest

The authors declare that they have no known competing financial interests or personal relationships that could have appeared to influence the work reported in this paper.

## Data availability

Data will be made available on request.

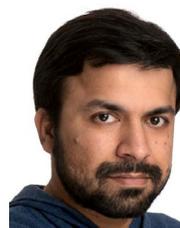

**Kashif Mehmood** received his B.Sc. in telecommunications engineering in 2012 from FASTNUCES, Islamabad, Pakistan, and, M.Eng. in information and communication engineering in 2019 from Sejong University, Seoul, South Korea. He is currently pursuing his Ph.D. in information and communication technology at Department of Information Security and Communication Technology at Norwegian University of Science and Technology (NTNU), Trondheim, Norway. He has worked as a researcher at Sejong University and various communication service providers throughout his professional career. His research interests include autonomous networks, game and information theory, model- and policy-based management and nextgeneration networking.

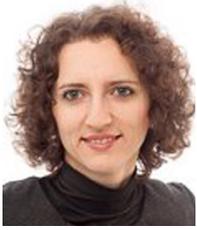

**Katina Kralevska** is an Associate Professor at the Department of Information Security and Communication Technology, NTNU (2018- ). She is also a co-founder of MemoScale, a company for data storage and compression. Her research interests and activities lie in the areas of network architectures and implementations; resource management in wireless networks and storage systems; coding theory and blockchain. She was a Deputy Head of Department for 2 years (2019-2020). She received her M.Sc. degree in mobile and wireless communications and Ph.D. degree in telematics in 2012 and 2016, respectively.

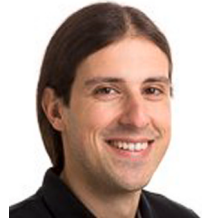

**David Palma** received the B.Sc. degree in informatics engineering in 2007 and a Ph.D. degree in information science and technology in 2013 from the University of Coimbra, Portugal. He is an Associate Professor at the Department of Information Security and Communication Technology, Norwegian University of Science and Technology (NTNU), Trondheim, Norway. He was an H2020 Marie Sklodowska-Curie Postdoctoral fellow at NTNU and has worked in the past as a Researcher and Project Manager at OneSource, Portugal, as well as an invited Assistant Professor at the University of Coimbra. His current research interests are on next-generation internet and computer networks.